# Hybrid Free-space-optics and Millimetre-wave D-band Transmitter enabled by Optically Harmonically Locked Lasers

Zichuan Zhou, Zun Htay, Amany Kassem, Izzat Darwazeh, Zhixin Liu

Department of Electronic and Electrical Engineering, University College London, zczlzz0@ucl.ac.uk

**Abstract** *We demonstrated hybrid free-space optics (FSO) and D-band (110-170GHz) millimetre wave transmitter enabled by a single phase-locked laser pair, simultaneously enabling ultra-low RF phase noise and optical linewidth for communications. Based on this, we further study combined capacity with beam angle misalignment using >100Gb/s signalling. ©2026 The Author(s)*

## Introduction

The increasing demand for capacity in satellite communication and wireless backhaul has been driving the development of high-capacity point-to-point wireless transmission [1-3]. Free-space optics (FSO) transceivers, as key enablers, have demonstrated 1-Tbps FSO transmission between observatories with large elevation differences, showing strong potential for ground-satellite feeder link [3] as well as direct wireless connection between radio units (RU) [2].

However, FSO transmission suffers from high loss in cloudy conditions, and it is inherently sensitive to angular misalignment due to its highly directional beam, thus requiring complex and costly pointing system [4]. Meanwhile, millimetre (mm) - wave, particularly D-band systems (110-170 GHz), are emerging as promising candidates for high-capacity wireless links, offering up to 60 GHz bandwidth and increasingly mature components/subsystems, such as D-band mixers, amplifiers and antennas [5]. Compared to FSO, mm-wave transmission link exhibit significantly higher tolerance to beam misalignment and lower loss in fog and cloud conditions, complementing FSO [6,7].

This has led to investigations of combining FSO and mm-wave transmission technologies for simultaneously achieving high capacity and resilient connection [7-12]. In [8,9], researchers have demonstrated hybrid FSO and Ka-band (centred at 28.5 GHz) transmission, achieving up to 400 Gbps and 24 Gbps data capacity for FSO and mm-wave link, respectively. Mm-wave signal is generated using electronic mixers, which modulate intermediate frequency (IF) signal onto an electronic multiplier-based local oscillator (LO) [8,9]. This, inherently, is a combining of two separate systems, where the mm-wave signal performance is inherently limited by the LO phase noise, which scales quadratically with frequency, limiting data capacity [13,14].

In the meantime, researchers have shown hybrid FSO and mm-wave transmission using photonics-assisted approach [10-12]. In [10], 160 Gbps FSO and 40 Gbps W-band transmission was demonstrated in urban environment using three free-running lasers at transmitter side, one was dedicated for FSO signal generation, and two were employed for mm-wave signal generation using optical heterodyne, which, inherently, is a combination of two independent photonic systems.

Although laser-based mm-wave generation has been employed for mm-wave connection [7, 10-12], most demonstration use free running lasers. This not only results in a drift of carrier frequencies at GHz-level, causing interference with signal transmitting at neighbouring frequency band [1], but also leading to high frequency and phase noise that reducing transmission capacity [1,13].

In this paper, we propose and demonstrate a hybrid FSO and mm-wave transmitter addressing the aforementioned problems, including generation of a precise and stable mm-wave carrier frequency, significantly improved mm-wave phase noise and optical linewidth, and generation of mm-wave and optical modulated channels using the same laser system [15]. This is achieved by our optical harmonic locking method that enables phase locking of two widely separated, Hz-level linewidth lasers, which simultaneously feed both optical and D-band modulators [13,15].

Based on this novel transmitter system, we demonstrate 17GBd 64QAM (102Gb/s data rate) transmission for both FSO and D-band signalling over a proof-of-concept 55-cm link. Further, to understand the system resilience that facilitating mm-wave-assisted laser beam pointing [16], we study transmission performance with regard to angular misalignment tolerance, where digital coherent combine is investigated to enhance the performance of the proposed hybrid transmitter.

## Experimental Setup

Fig. 1a shows the experimental setup for hybrid FSO and mm-wave D-band transmission system. The modulation signal was generated using a 65GSa/s arbitrary waveform generator (AWG) with 25 GHz bandwidth and 8-bit digital resolution. A pseudorandom binary sequence (PRBS) of $2^{20}$-1 length was used to generate 17-GBd, 16 and 64 QAM digital signals shaped by root-raised

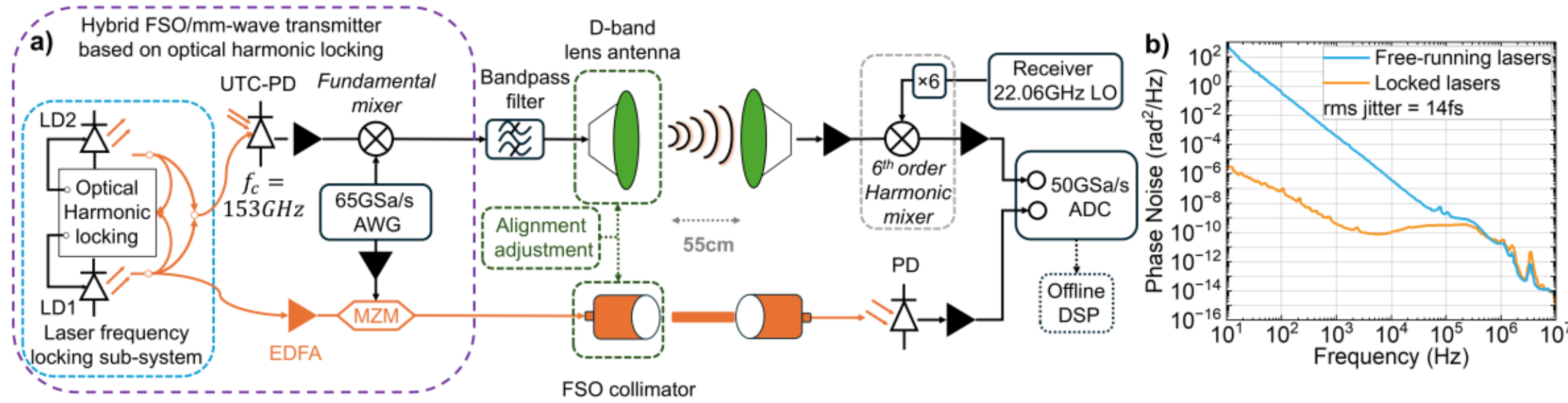

**Fig. 1:** a) Experimental setup for combined FSO and mm-wave D-band transmission system; b) Phase noise of carrier generated by beating two free-running lasers (blue) and two frequency and phase locked lasers (orange)

cosine (RRC) pulse filter with a roll-off factor of 0.1, resulting in a baseband signal of 8.5 GHz bandwidth. The baseband signal was digitally up-converted to a carrier centred at 11.5 GHz, resulting in a real-valued doubled sideband sub-carrier-modulated (SCM) signal over 3-20 GHz.

At transmitter side, two ultra-low linewidth lasers with sub-100-Hz linewidth emitting 15dBm were used. For mm-wave D-band signal generation, the frequency spacing between LD1 and LD2 was adjusted to 153 GHz. The frequency and phase of LD1 and LD2 were locked using optical harmonic locking technique [1,13,15] and then combined and fed into a D-band UTC-PD, outputting a 153 GHz carrier with around -8dBm power, which was further amplified by a 20-dB gain D-band amplifier. The measured phase noise of 153 GHz carrier is shown by Fig. 1b. Compared to free-running lasers case (blue curve), our optical harmonic locking technique (orange curve) significantly improves phase noise in low frequency region. For example, at 10Hz offset, the signal generated by the locked laser pair yields at single side band (SSB) phase noise of 2.3e-6 rad$^2$/Hz, whilst the unlocked laser pair was measured 791 rad$^2$/Hz. This indicates a precise and stable frequency generated in our transmitter.

The rms jitter (integrated from 10Hz to 10MHz) of generated 153 GHz carrier is 14fs [13,15], enabling up to 15dB receiver sensitivity improvement compared to conventional electronic oscillators as demonstrated in [13]. The 153GHz carrier was then fed into D-band fundamental mixer for frequency up-conversion. The upconverted double sideband mm-wave signal was fed into a 133-150 GHz D-band bandpass filter, resulting in >20dB suppression of upper sideband. The output of bandpass filter (around -29dBm) was fed into a D-band lens-antenna with 32. 5dBi gain and received by an identical lens-antenna pair spaced 55cm away. The maximum received power measured at antenna output was around -35dBm, corresponding to a total of 6dB D-band link loss. The receiver antenna output was connected to a D-band variable attenuator, followed by a 21-dB gain D-band amplifier with 6.5 dB noise figure, and a 6$^{th}$ order harmonic mixer-based frequency down-converter. The resulting baseband signal was detected by a 50GSa/s 8-bit real-time oscilloscope.

For FSO transmission, an erbium-doped fibre amplifier (EDFA) amplified the tapped LD1 output to 15dBm, which was then fed into a Mach-Zehnder modulator (MZM) biased at quadrature point, driven by the AWG. The MZM output was fed into the transmitter side collimator, and the optical beam was collected by the same model collimator, placed 55cm away from transmitter. The maximum received optical power was around 7 dBm, corresponding to 2 dB FSO link loss, mainly from optical coupling imperfections. The collected beam was detected using a 40 GHz photodetector, followed by a 25 dB gain RF amplifier before entering the 50GSa/s 8-bit real-time oscilloscope.

For both FSO and mm-wave link, the receiver DSP involves digital down-conversion, matched filtering, clock and frame synchronization, and a digital equalizer [17] with 131 first order taps and 3 third-order taps. To study misalignment tolerance, the collimator and lens-antenna were mounted on two separate adjustment stages, allowing up to 6.4- and 4-degree azimuth and elevation adjustment. To investigate SNR improvement using hybrid scheme with coherent combine detection [18], same data was transmitted on both FSO and D-band channels, the delay difference of two channels was calculated offline, and the synchronized samples were fed into a 2 by 1 digital equalizer for coherent combination [17].

## Experimental Results

Fig. 2a and b show the measured BER at different received power for 17GBd 16/64QAM FSO and D-band signal respectively. D-band power was measured at receiver antenna output and FSO power was measured at photodetector input. For D-band signal, the HD-FEC (BER=3.8e-3 [19]) receiver sensitivity for 16QAM is around -45dBm. The SD-FEC (BER=2e-2 [20]) receiver sensitivity for 64QAM is around -43dBm. For FSO signal, the HD-FEC receiver sensitivity for 16QAM is around -6.5dBm. The SD-FEC receiver sensitivity for 64QAM is around -5dBm. Considering SD-FEC BER, the link budget for

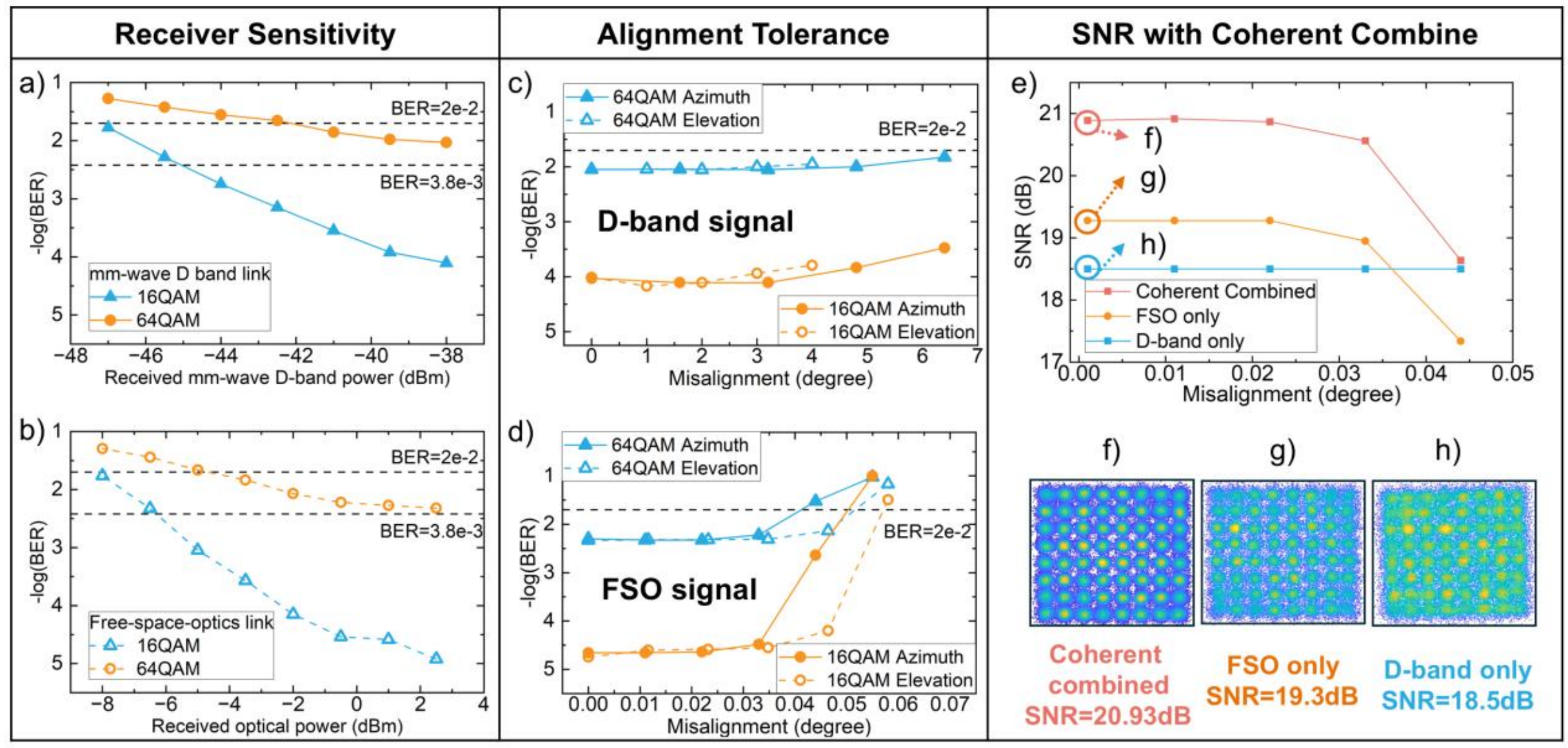

**Fig. 2:** BER vs received power for a) D band link and b) FSO link, BER with azimuth and elevation misalignment for c) D-band and d) FSO signal, e) SNR comparison of different scheme, constellation diagram of f) hybrid, g) FSO only and h) D-band only

64QAM D-band and FSO signal are both 14dB. For 64QAM, at maximum received power, D-band signal exhibits slightly worse performance (BER=1e-2) than FSO signal (BER=5e-3) due to high conversion loss of down-convertor.

Fig. 2c and d show the measured BER as a function of azimuth and elevation misalignment for D-band and FSO signal, respectively. For the D-band signal (Fig. 2c), no measurable BER degradation is observed for misalignment angles up to 3°, beyond which the BER only slightly increased from 9.5e-5 to 3.4e-4 for 16QAM and 9e-3 to 1.5e-2 for 64QAM, respectively, which remain below the SD-FEC threshold. Similar trend was observed for the elevation misalignment. For D-band signal, the maximum misalignment angle is tested in this work limited by our alignment stage. For FSO signals, the link exhibits a significantly higher sensitivity to beam misalignment, with the BER exceeding SD-FEC threshold when azimuth and elevation misalignment angles as small as 0.05° for both 16 and 64QAM. This substantial difference clearly show that D-band signal offers more than 128× (6.4° vs. 0.05°) and 80× (4° vs. 0.05°) greater misalignment tolerance than FSO signal. This feature of hybrid transmitter can potentially reduce pointing system complexity using mm-wave signal as beacon signal for coarse alignment [16].

Finally, Fig. 2e shows the 64QAM SNR at different azimuth misalignment under three different scenarios: D-band only (blue marker), FSO only (orange marker) and hybrid FSO/D-band with digital coherent combine (red marker). When transmitter and receiver are perfectly aligned, the FSO, D-band and hybrid (FSO/D-band with digital coherent combine) signal SNR are 19.3, 18.5 and 20.93dB, respectively, with constellation diagrams at back-the-back shown in Fig. 2f-g. This indicates that gain 1.6 dB and 2.4 dB SNR improvement using hybrid solution compared to FSO only and D-band only cases. When the azimuth misalignment angle increases, the SNR improvement offered by the hybrid transmitter diminishes. Nevertheless, under well-aligned conditions, the hybrid FSO/D-band achieves higher SNR than any link along, supporting higher capacity transmission over longer distance. Furthermore, even when the FSO link is severely impaired by misalignment, the hybrid architecture maintains BER below the SD-FEC threshold, demonstrating the resilience benefit of coherent signal combining.

## Conclusion

We proposed and demonstrated a novel hybrid FSO/D-band mm-wave transmitter using optical harmonic locking technique, enabling frequency stable and low phase noise mm-wave carrier generation as well as high quality FSO and D-band data signal transmission. In the proof-of-concept, we demonstrated 17GBd 64QAM FSO and D-band transmission, achieving 102Gbps per channel. We show that D-band link offers more than 80 times greater tolerance to angle misalignment and demonstrate 2.4dB SNR improvement by digital combining D-band and FSO, showing potential for D-band assisted FSO pointing with enhanced system capacity and resilience.

## Acknowledge

The authors would like to acknowledge funding from the EPSRC grant TRACCS(EP/W026732/1), PROSPECT(EP/Z534444/1) and EU-horizon grant 6G-MUSICAL(101139176).